\def\lesssim{\mathrel{\hbox{\rlap{\hbox{\lower4pt\hbox{$\sim$}}}\hbox{$<$}}}}

\def\gtrsim{\mathrel{\hbox{\rlap{\hbox{\lower4pt\hbox{$\sim$}}}\hbox{$>$}}}}

\def\ll_lsun{log$({L/\rm L_{\odot}})$~}

\def\masa_msun{$M/ \rm M_{\odot}$~}

\def\m_mstar{$M/M_{*}$~}

\documentclass{aa}
\usepackage{graphicx}
        
\begin{document}

\title{The  formation of DA white dwarfs with thin hydrogen envelopes}

\author{L. G. Althaus$^1$\thanks{Member of the Carrera del Investigador
        Cient\'{\i}fico y Tecnol\'ogico  and IALP, CONICET / FCAG-UNLP, 
        Argentina.},
        M. M. Miller Bertolami$^2$\thanks{Fellow of CONICET, Argentina},
        A.  H.  C\'orsico$^{2,3\star}$,
        E. Garc\'{\i}a-Berro$^{1,4}$, \\ and
        P. Gil-Pons$^1$ }
\offprints{L. G. Althaus}

\institute{
$^1$ Departament de F\'\i  sica Aplicada, Universitat Polit\`ecnica de
Catalunya,  Av.  del  Canal  Ol\'\i mpic,  s/n, 08860,  Castelldefels,
Barcelona,  Spain\\   $^2$  Facultad  de   Ciencias  Astron\'omicas  y
Geof\'{\i}sicas, Universidad  Nacional de  La Plata, Paseo  del Bosque
S/N,   (B1900FWA)   La   Plata,   Argentina.\\   $^3$   Instituto   de
Astrof\'{\i}sica   La  Plata,   IALP,  CONICET-UNLP\\   $^4$  Institut
d'Estudis Espacials de Catalunya, Ed. Nexus, c/Gran Capit\`a 2, 08034,
Barcelona, Spain.\\
\email{althaus@fcaglp.unlp.edu.ar} }

\date{Received; accepted}

\abstract{We study  the  formation  and evolution of  DA white dwarfs,
the progenitors  of which have  experienced a late thermal  pulse (LTP)
shortly after  the departure from  the thermally pulsing AGB.  To this
end,  we  compute the  complete  evolution  of  an initially  $2.7  \,
M_{\sun}$  star all the  way from  the zero-age  main sequence  to the
white dwarf stage.  We find that most of  the original H-rich material
of the post-AGB  remnant is burnt during the  post-LTP evolution, with
the  result that,  at  entering  its white  dwarf  cooling track,  the
remaining H  envelope becomes $10^{-6} \, M_{\sun}$  in agreement with
asteroseismological inferences for some ZZ Ceti stars.
\keywords{stars:  evolution   ---   stars: abundances ---  stars:  AGB
stars: interiors --- stars: white dwarfs --- stars: oscillations } }

\authorrunning{Althaus et al.}

\titlerunning{The  formation of DA white dwarfs with thin H   
envelopes.}

\maketitle


\section{Introduction}

White-dwarf stars  constitute the end-point  of the evolution  of low-
and intermediate-mass stars. Hence, they  play a key role in our quest
for understanding the  structure and history of our  Galaxy.  The vast
majority of white  dwarfs are the remnants of  Asymptotic Giant Branch
(AGB)  stars  and are  characterized  by  H-rich atmospheres  (usually
referred  to as  DAs).  Indeed,  white  dwarfs with  H surface  layers
represent about 80\% of the spectroscopically identified white dwarfs.

Considerable progress  in the study  of these stars has  been possible
partially  because some  of these  stars  pulsate.  The  study of  the
pulsational  pattern of variable  DA white  dwarfs (or  ZZ~Ceti stars)
through  asteroseismology has provided  valuable constraints  to their
fundamental properties, such as  the core composition, the outer layer
chemical stratification  or the stellar mass. In  particular, the mass
of  the outer H  layer is  an important  issue regarding  the spectral
evolution  theory.   Most   of  the  asteroseismological  fittings  to
individual  pulsators  predict  the  thickness of the almost pure H envelope 
to be about
$10^{-4}$  of the  total stellar  mass, $M_*$,  in agreement  with the
expectations from  the standard stellar evolution  theory.  However, H
envelopes substantially  thinner than the quoted  value have been
inferred for some ZZ~Ceti  stars (Bradley 2001), suggesting that there
is likely a  {\sl range} of H content possible for  DA white dwarfs, a
suggestion that  is in line with evidence  from spectroscopy (Fontaine
\& Wesemael 1997).  The existence of  DA white dwarfs with such thin H
envelopes  is difficult  to reconcile  with the  accepted view  of the
post-AGB evolution theory  that predicts the thickness of the H envelope
to be  about $10^{-4} M_*$.  Indeed, the  existence of such white
dwarfs poses a real challenge to the theory of stellar evolution.

\begin{figure}
\centering
\includegraphics[clip,width=250pt]{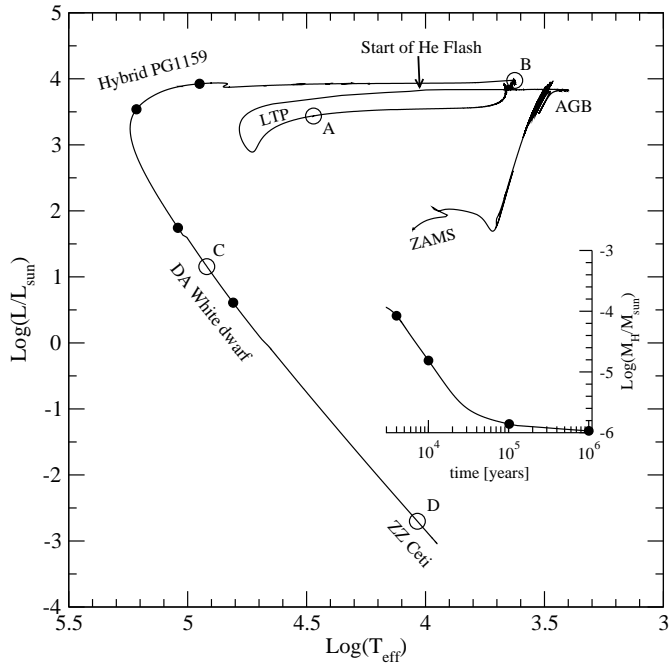}
\caption{H-R diagram for the complete evolution  of the initially $2.7
\, M_{\sun}$  stellar  model  from  the  ZAMS  to  the  ZZ~Ceti domain.   
In this simulation, the star undergoes a LTP shortly after leaving the
AGB at $\log T_{\rm eff}= 4$.  As a result, further evolution proceeds
to the blue  and then to the red-giant domain where  the H envelope is
diluted by surface convection. Inset:  temporal evolution of H mass in
solar units.   Filled dots correspond to those  stages indicated along
the evolutionary  track.  Most  of H is  processed by  nuclear burning
during the PG~1159 stage.}
\label{franky.eps}
\end{figure}

We undertake the  current investigation with the aim  of exploring the
possibility that a fraction of the DA white dwarf population with thin
H envelopes  could be the  descendants of post-AGB PG~1159  stars with
surface layers  rich in  H, helium, carbon  and oxygen  (Werner 2001).
According  to  Bl\"ocker (2001),  these  hybrid-PG~1159  could be  the
result of  a final  helium shell flash  experienced by a  star shortly
after the departure  from the AGB --- the  so-called AGB final thermal
pulse, AFTP scenario,  or late thermal pulse, LTP.   During an AFTP or
LTP, H is  not burnt, in contrast to post-AGB  stars that experience a
{\sl very}  late thermal pulse  (a born-again episode, see  Althaus et
al.  2005 for  a recent  calculation), but  it is  diluted  by surface
convection  and mixed  inwards with  the underlying  intershell region
formerly enriched in  helium, carbon and oxygen.  We  speculate that H
burning during the  following PG~1159 stage will reduce  most of the H
content   of   the   remnant   to   a  level   compatible   with   the
asteroseismological  predictions   for  some  DA   white  dwarfs  with
relatively  thin  H  envelopes.   Specifically,  we  present  complete
evolutionary calculations that simulate the formation and evolution of
H-rich white  dwarfs through a  late helium shell flash  starting from
the main sequence.
 

\section{Input physics and computational details}

The calculations presented  in this work have been  done with the same
stellar evolutionary  code we  employed in our  previous study  of the
formation  of  H-deficient  white  dwarfs  via  a  born-again  episode
(Althaus et al. 2005). The code  is based on a detailed description of
the  main physical  processes involved  in the  formation  of post-AGB
stars. Abundance  changes are described  by means of  a time-dependent
scheme for the simultaneous  treatment of nuclear evolution and mixing
processes (Althaus et al. 2003).  This is particularly relevant during
some  short-lived  phases of  evolution  for  which the  instantaneous
mixing  approximation  becomes   inadequate  for  addressing  chemical
mixing.  In addition, we have included time-dependent overshoot mixing
above and below any  convective region during all evolutionary stages.
As shown  by Herwig (2001),  convective overshoot during  the post-AGB
stage is required to obtain  efficient dregde-up for very low envelope
masses  and dilution  of H\footnote{An  additional process  that could
contribute to  the dredge-up is  the rotationally induced  mixing, see
Langer et  al. (1999)}.  Gravitational settling, chemical  and thermal
diffusion have  been considered during  the whole white  dwarf regime.
The  evolutionary stages  for  the progenitor  star  are described  in
Althaus et al.   (2005).  Briefly, the evolution of  an initially $2.7
\, M_{\sun}$  stellar model has  been computed from the  zero-age main
sequence all  the way from the stages  of H and helium  burning in the
core up  to the tip of the  AGB where helium thermal  pulses occur.  A
solar-like initial composition $(Y,Z)= (0.275,0.02)$ has been adopted.
After experiencing 10 thermal  pulses, the progenitor departs from the
AGB with  a stellar mass of  $0.5885 \, M_{\sun}$  and evolves towards
high effective temperature ($T_{\rm eff}$) values.  Departure from the
AGB takes  place at such an  advanced phase of the  helium shell flash
cycle  that the  post-AGB remnant  undergoes  a LTP  at about  $T_{\rm
eff}$=  10000~K. In order  to get  the last  flash at  the appropriate
location, we forced the model to depart from the thermally pulsing AGB
somewhat later than it was done in Althaus et al. (2005).  As a result
of the flash,  the remnant evolves back to the  AGB, and eventually to
the domain of the PG~1159 stars.

\begin{figure}
\centering
\includegraphics[clip,width=270pt]{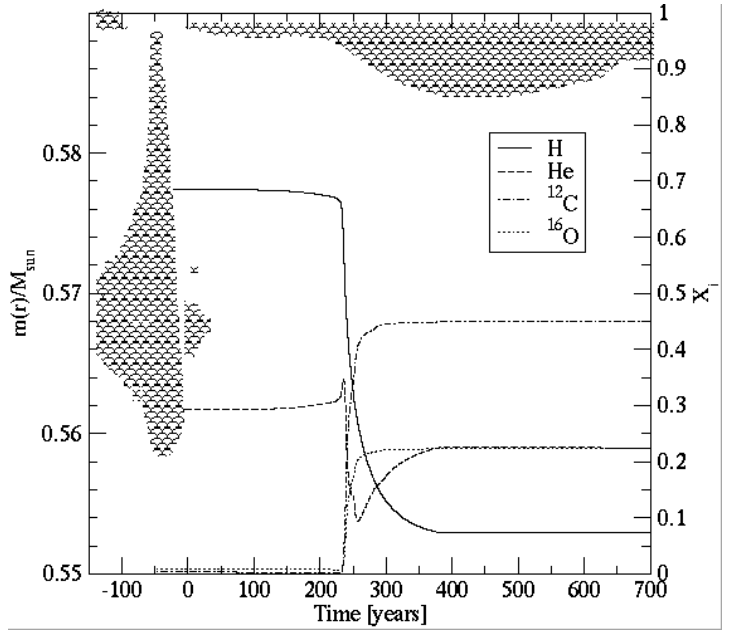}
\caption{Temporal  evolution  of  the  convective  regions and surface 
abundances  of $^{1}$H,  $^{4}$He,  $^{12}$C and  $^{16}$O during  the
LTP.  Note the  development of  the  surface convection  zone and  the
dilution of surface H.  }
\label{conve.eps}
\end{figure}

\begin{figure*}
\centering
\includegraphics[clip,width=450pt]{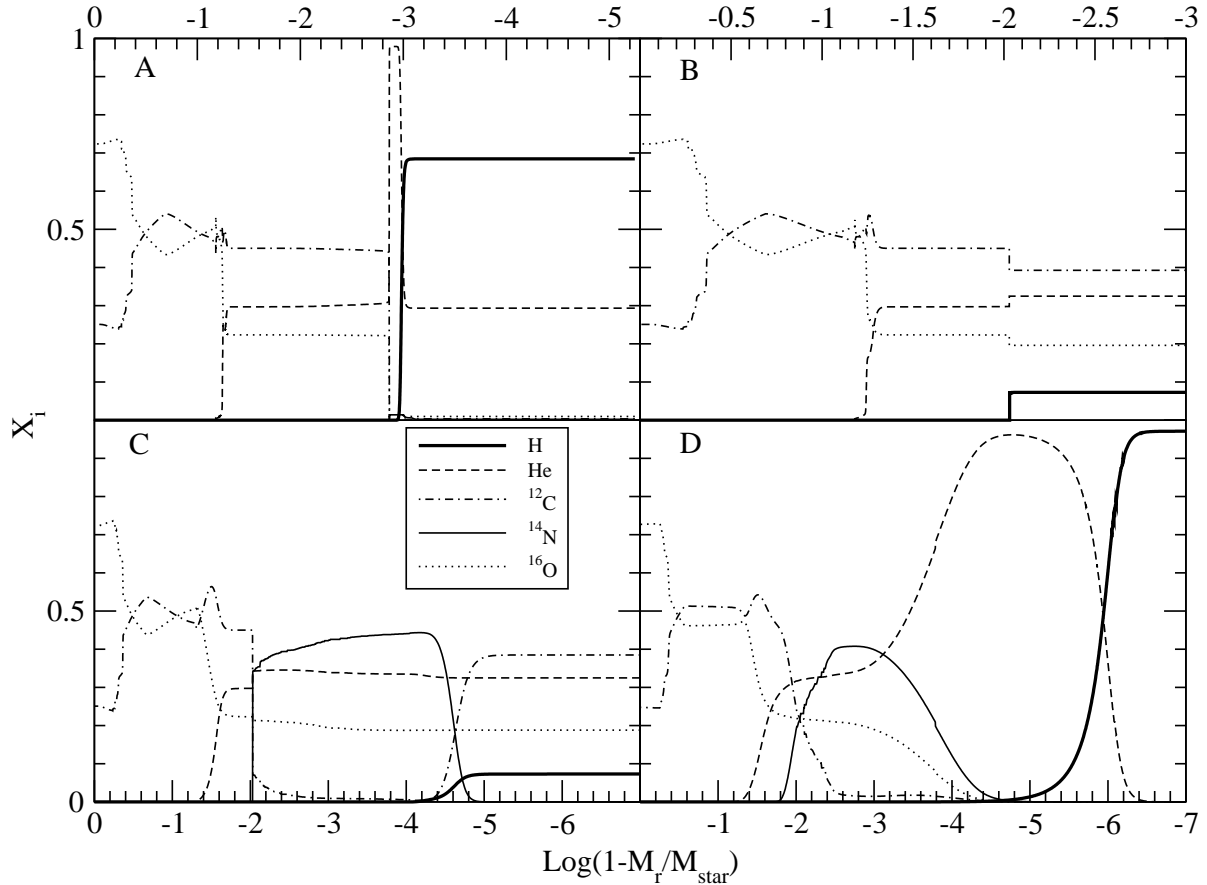}
\caption{Internal  abundance   distribution   of  $^{1}$H,   $^{4}$He, 
$^{12}$C,  $^{14}$N and  $^{16}$O  as  a function  of  the outer  mass
fraction  for the  $0.5885 \,  M_{\sun}$ remnant  at  various selected
epochs comprising the LTP and the  white dwarf stages.  Panels A, B, C
and  D correspond  to stages  labeled by  letters along  the  curve in
Fig.~1. Because of nuclear  burning occurring during the PG~1159 stage
at high effective temperatures, the  remnant star is left with a small
H  content at entering  the white  dwarf cooling  track. Gravitational
settling ultimately leads to the formation  of a DA white dwarf with a
thin H envelope.}
\label{quimi_nuevo.eps}
\end{figure*}


\section{Evolutionary results}

A complete coverage  of the evolutionary stages in  the H-R diagram of
our sequence is  displayed in Fig.~\ref{franky.eps}. Relevant episodes
are shown  at specific  points along the  curve.  In  particular, note
that after the occurrence of the thermal pulses at the tip of the AGB,
the remnant  star evolves through a  LTP, which starts  at about $\log
T_{\rm  eff} =4$.   As  a result,  the  star evolves  rapidly to  high
$T_{\rm eff}$ values and then  back to the red-giant region, where the
remaining H-rich  material is diluted by surface  convection and mixed
inwards with the underlying intershell region (see Fig. 2).  After the
short-lived dredge-up  episodes, the remnant is left  with an envelope
composition rich in H, helium, carbon and oxygen.  In particular, H is
diluted to  very low  abundances and it  extends downwards as  deep as
$5.6 \times 10^{-3} \, M_{\sun}$ below the stellar surface.  Dredge-up
sets  in  at the  minimum  effective  temperature ($T_{\rm  eff}\simeq
4\,600$~K)     and     develops     during     a     short     redward
excursion. Specifically,  the final surface  abundances by mass  of H,
helium,  carbon and  oxygen  amounts, respectively,  to 0.073,  0.323,
0.382 and 0.19.  After H  dilution, evolution proceeds into the domain
of the  central stars  of planetary nebulae  at high $T_{\rm  eff}$ to
become a  hybrid PG~1159  star.  In the  meantime, H is  reignited and
after the  point of  maximum effective temperature  is reached  on the
early white dwarf cooling track, the  H content is reduced to about $8
\times  10^{-7} \,  M_{\sun}$, that  is, by  more than  two  orders of
magnitude as  compared with  the amount  of H that  is left  after the
remnant departs from the AGB. This value is the amount of H with which
the star enters the white  dwarf domain. The temporal evolution of the
H content is illustrated in the inset of Fig.  \ref{franky.eps}.  Most
of  the  residual  H  material  is  burnt over  a  period  of  roughly
100\,000~yr during the PG~1159 stage.  It is noteworthy that H burning
takes place in a  carbon-enriched medium, with the consequent nitrogen
enrichment in the H-burning layers.   On the cooling track, H diffuses
outwards, turning the white dwarf into  one of the DA type with a thin
pure H envelope of a few $10^{-7} \, M_{*}$.

An  example of  the main  inner chemistry  variations that  take place
after   the   remnant   star   leaves   the   AGB   is   provided   in
Fig. \ref{quimi_nuevo.eps}.   Specifically, the abundances  by mass of
$^{1}$H, $^{4}$He,  $^{12}$C, $^{14}$N  and $^{16}$O are  displayed in
terms  of  the  outer  mass  fraction.  Panel  A  shows  the  chemical
stratification shortly after the occurrence of the LTP (marked as A in
Fig.~1).   In the  outermost layers  the chemical  profile corresponds
essentially  to that emerging  from the  dredge-up events  and nuclear
burning during  the thermally pulsing  AGB phase. As the  star becomes
cooler,  the inward-growing surface  convection zone  eventually mixes
downwards  protons   from  the  original  H-rich   envelope  with  the
underlying intershell  region rich in helium, carbon  and oxygen.  The
resulting  abundance  distribution  after  this dredge-up  episode  is
illustrated in panel B (corresponding to point B in Fig.~1). Note that
H has been diluted to surface abundances of about 0.073 and mixed down
to  regions as  deep as  $5.6 \times  10^{-3} \,  M_{\sun}$  below the
stellar  surface. Most  of  this deep-lying  H  is burnt  as the  star
evolves towards high effective  temperatures, as documented in panel C
(labeled as C in Fig.~1).   After nuclear processing is completed, the
remaining  H mass  amounts to  only  $8 \times  10^{-7} \,  M_{\sun}$.
Because H burning occurs  in a carbon-enriched medium, an overwhelming
abundance  of   nitrogen  is   expected.   Finally,  panel   D  (which
corresponds to  point D  in Fig.~1) illustrates  the situation  by the
time the domain of the ZZ  Ceti stars is reached.  The signatures left
by the various diffusion processes acting during white dwarf evolution
are easily recognizable.  In particular,  note the formation of a pure
H  envelope  of  $4  \times  10^{-7}  \,  M_{\sun}$  as  a  result  of
gravitational  settling. That  is,  a DA  white  dwarf with  a thin  H
envelope is formed.  Gravitationally-induced diffusion gives also rise
to   the   development   of   a  triple-layered   chemical   structure
characterized by  a pure H envelope  atop a pure helium  buffer and an
underlying intermediate remnant shell rich in helium, carbon, nitrogen
and  oxygen. Note  also the  mixing episode  that takes  place  in the
region below the intershell zone around $1-M_r/M_*= 0.1$.  This region
is characterized by a  inward-decreasing mean molecular weight induced
by  the occurrence of  overshooting during  the AGB  thermally pulsing
phase.  The resulting  Rayleigh-Taylor rehomogenization is responsible
for the redistribution of the chemical species in that region.


\section{Discussion and conclusions}

The  present  work  has  been  specifically designed  to  explore  the
possibility that post-AGB  stars that undergo a LTP  could evolve into
the white dwarf  regime with thin H envelopes.   The results are based
on evolutionary calculations  starting on the ZAMS that  cover all the
stages involved in the formation  of hybrid (H-rich) PG~1159 stars via
a LTP.  Our results show that  most of the original H-rich material of
the post-AGB remnant is burnt  during the post-LTP evolution.  We have
found  that when  the star  enters  its terminal  white dwarf  cooling
track, the H content is reduced  to about $ 10^{-6} \, M_{\sun}$ after
nuclear burning has virtually ceased, that is, by more than two orders
of magnitude as  compared with the amount of H that  is left after the
remnant departs  from the  AGB. Because we  have not  invoked possible
mass-loss episodes during the PG~1159  stage, the quoted value for the
H mass should be considered as an upper limit.

Roughly 25\%  of stars leaving the  AGB are expected to  suffer from a
born-again episode  (with the  complete burning of  protons) or  a LTP
when  H burning  is  still  active (Bl\"ocker  2001;  Iben 1984).   In
particular,  stars leaving the  AGB with  a thermal-pulse  cycle phase
larger than about 0.90 are expected  to undergo a LTP.  Hence, in view
of our  results, it is  conceivable that a non-negligible  fraction of
the DA  white dwarf population  could harbour thin H  envelopes.  This
result   is  in  line   with  asteroseismological   and  spectroscopic
predictions  that suggest  that there  is likely  a range  of  H thicknesses
 possible  for DA  white  dwarfs  (Fontaine  \& Wesemael  1997;
Bradley 1998,  2001).  Indeed, H envelopes  substantially thinner
than the  canonical value of $10^{-4}  \, M_*$ have  been inferred for
some ZZ~Ceti  stars.  In particular, Bradley \&  Kleinman (1997) found
from seismological data a H  envelope thickness of $5 \times 10^{-7}\, M_*$
for the ZZ~Ceti star G~29-38.  In addition, seismologically-inferred H
envelopes  ranging from $\sim  10^{-6}$ to  $\sim 10^{-7}\,  M_*$ have
been reported by Bradley (1998) for G~117-B15A and R~548 if the period
of the  215~s mode  corresponds to an  $\ell= 1$, $k=1$  mode. Coupled
with previous  inferences, Bradley (1998)  concluded that the  H layer
masses of several  ZZ Ceti pulsators should lie  between $10^{-4}$ and
$10^{-7}\,  M_*$.   The  results  presented  here  place  on  a  solid
evolutionary basis  these predictions about  the existence of  some DA
white dwarfs with thin H envelopes.


\begin{acknowledgements}
L.G.A acknowledges the Spanish MCYT  for a Ram\'on y Cajal Fellowship.
We also acknowledge to our referee, A. Weiss, for his report.  Part of
this work has  been supported by the Instituto  de Astrof\'{\i}sica La
Plata, by the  MCYT grant AYA2002--4094--C03--01, by the  CIRIT and by
the European Union FEDER funds.
\end{acknowledgements}

\end{document}